\documentclass[onecolumn,letter]{IEEEtran}
\usepackage{amsmath}
\usepackage{latexsym}
\usepackage{latexsym}
\usepackage{amssymb}
\usepackage{bm}
\usepackage[dvips]{graphicx}

\newtheorem{thm}    {Theorem}%[section]
\newtheorem{lem}     {Lemma}%[section]
\newtheorem{cor}  {Corollary}%[section]
\newtheorem{rem}     {Remark}%[section]
\def\argmax{\mathop{\rm argmax}}
\def\argmin{\mathop{\rm argmin}}

\def\vlimsup{\varlimsup}

\def\Label#1{\label{#1}\ [\ #1\ ]\ }
\def\Label{\label}

\def\Tr{\mathop{\rm Tr}\nolimits}

\begin{document}
\title
{Discrimination of two channels by adaptive methods and its application to quantum system}
\author{
Masahito Hayashi
\thanks{
M. Hayashi is with Graduate School of Information Sciences, Tohoku University, Aoba-ku, Sendai, 980-8579, Japan
(e-mail: hayashi@math.is.tohoku.ac.jp)
}}
\date{}
\maketitle

\begin{abstract}
The optimal exponential error rate for adaptive discrimination of two channels is discussed.
In this problem, adaptive choice of input signal is allowed.
This problem is discussed in various settings.
It is proved that 
adaptive choice does not improve the exponential error rate in these settings.
These results are applied to quantum state discrimination.
\end{abstract}

\begin{keywords}
%Keywords:
Simple hypothesis testing,
Channel,
Discrimination,
Quantum state,
One-way LOCC,
Active learning,
Experimental design,
Stein's lemma,
Chernoff bound,
Hoeffding bound,
Han-Kobayashi bound
\end{keywords}

\section{Introduction}
\PARstart{D}{iscriminating} two distributions is
treated as a fundamental problem in the field of statistical inference.
This problem can be regarded as 
simple hypothesis testing because both hypotheses consist of 
a single distribution.
Many researchers, Stein, Chernoff\cite{Chernoff}, Hoeffding\cite{Hoe}, and Han-Kobayashi\cite{HK} 
have studied the asymptotic behavior when the number $n$ of identical and independent observations is sufficiently large.
They formulated a simple hypothesis testing/discrimination of two distributions
 as an optimization problem and derived the respective optimum value, e.g., 
the optimal exponential error rate.	
We call these optimum values the Stein bound, the Chernoff bound, the Hoeffding bound,
and the Han-Kobayashi bound, respectively.
Han \cite{Han-test,Han} later extended these results to the discrimination of two general sequences of distributions, including the Markovian case.
Nagaoka-Hayashi \cite{N-H} simplified Han's discussion and generalized Han's extension of the  Han-Kobayashi bound.

In the present paper, we consider another extension of the above results.
That is, we extend the above results to the discrimination of two (classical) channels,
in which two probabilistic transition matrices are given.
Such a problem has appeared in Blahut\cite{blahut}.
In this problem, 
the number of applications of this channel is fixed to a given constant $n$, and 
we can choose appropriate inputs for this purpose.
In this case, we assume that the given channel is memoryless.
If we use the same input to all applications of the given channel,
the $n$ output data obeys an identical and independent distribution.
This property holds even if we choose the input randomly based on the same distribution on input signals.
This strategy is called the {\it non-adaptive} method.
In particular, when the same input is applied to all channels, it is called the {\it deterministic non-adaptive} method.
If the input is determined stochastically, it is called the {\it stochastic non-adaptive} method,
which was treated by Blahut\cite{blahut}.
In the non-adaptive method, our task is choosing the optimal input
for distinguishing two channels most efficiently.
In the present paper, we assume that we can choose the $k$-th input signal based on the preceding $k-1$ output data.
This strategy is called the {\it adaptive method}, which is the main focus of the present paper.
In the parameter estimation, such an adaptive method improves estimation performance.
That is, in the one-parameter estimation, the asymptotic estimation error is bounded by the inverse of the optimum Fisher information. However, if we do not apply the adaptive method, it is generally impossible to realize the optimum Fisher information in all points at the same time.
It is known that the adaptive method realizes the optimum Fisher information in all points\cite{HMadaptive,Fujiwara}.
Therefore, one may expect that the adaptive method improves the performance of discriminating two channels.

As our main result, we succeeded in proving that the adaptive method cannot improve the non-adaptive method in the sense of all of the above mentioned bounds, i.e.,
the Stein bound, the Chernoff bound, the Hoeffding bound, and the Han-Kobayashi bound.
That is, there is no difference between the non-adaptive method and the adaptive method in these asymptotic formulations.
Indeed, as is proven herein, the deterministic non-adaptive method gives the optimum performance with respect to the Stein bound, the Chernoff bound, and the Hoeffding bound.
However, in order to attain the Han-Kobayashi bound, in general, we need the stochastic non-adaptive method.

On the other hand, the research field in quantum information has treated the discrimination of two quantum states.
Hiai-Petz\cite{HP} and Ogawa-Nagaoka\cite{Oga-Nag:test} proved the quantum version of Stein's lemma.
Audenaert et al. \cite{spain} and Nussbaum-Szko{\l}a \cite{szkola,nussbaum} obtained the quantum version of the Chernoff bound.

Ogawa-Hayashi \cite{Oga-Hay} derived a lower bound of the quantum version of the Hoeffding bound.
Later, Hayashi \cite{hayashi06} and Nagaoka \cite{nagaoka06} obtained its tight bound
based on the results by Audenaert et al. \cite{spain} and Nussbaum-Szko{\l}a \cite{szkola,nussbaum}.
Hayashi \cite{Hayashi} (in p.90) obtained the quantum version of the Han-Kobayashi bound based on Nagaoka\cite{Naga-EQIS}'s discussion.
These discussions assume that any measurement on the $n$-tensor product system is allowed for testing the given state.
Hence, the next goal is the derivation of these bounds under some locality restrictions on an $n$-partite system for possible measurements.
One easy setting is restricting the present measurement to be identical to that in the respective system.
In this case, our task is the choice of the optimal measurement on the single system.
By considering the measurement and the quantum state as the input and the channel, respectively, we can treat this problem by the non-adaptive method of the classical channel.
Another setting is restricting our measurement to one-way local operations and classical communications (one-way LOCC).
In the above-mentioned correspondence, the one-way LOCC setting can be regarded as the adaptive method of the classical channel.
Hence, applying the above argument to discrimination of two quantum states, we can conclude that one-way communication does not improve discrimination of two quantum states in the respective asymptotic formulations.

Furthermore, the same problem appears in adaptive experimental design and active learning.
In learning theory, we identify the given system by using the obtained sequence of input and output pairs. 
In particular, in active learning, we can choose the inputs using the preceding data.
Hence, the present result indicates that active learning does not improve the performance of learning when the candidates of the unknown system are given by only two classical channels.
In experimental design, we choose suitable design of our experiment for inferring the unknown parameter.
Adaptive improvement for the design is allowed in adaptive experimental design.
When the candidates of the unknown parameter are only two values,
the obtained result can be applied.  
That is, adaptive improvement for design does not work.

The remainder of the present paper is organized as follows.
Section \ref{s2} reviews the Stein bound, the Chernoff bound, the Hoeffding bound, and the Han-Kobayashi bound in discrimination of two probability distributions.
In Section \ref{s3}, we present our formulation and notations of the adaptive method in the discrimination of two (classical) channels, and discuss the adaptive-method versions of the Stein bound, the Chernoff bound, the Hoeffding bound, and the Han-Kobayashi bound, respectively.
In Section \ref{s4}, we consider a simple example, in which
the stochastic non-adaptive method is required for
attaining the Han-Kobayashi bound.
In Section \ref{s5}, we apply the present result to
discrimination of two quantum states by one-way LOCC.
In Sections \ref{s6}, \ref{s7}, and \ref{s8}, we prove 
the adaptive-method versions of Stein bound, the Chernoff bound, the Hoeffding bound, and the Han-Kobayashi bound, respectively.

\section{Discrimination/simple hypothesis testing between two probability distributions}\Label{s2}
In preparation for the main topic,
we review the simple hypothesis testing problem for the
null hypothesis $H_0$ : $P^n$ versus the alternative hypothesis $H_1$: ${\overline{P}}^n$, 
where $P^n$ and ${\overline{P}}^n$ are the $n$-th identical and independent distributions of
$P$ and $\overline{P}$, respectively on the probability space ${\cal Y}$.
The problem is to decide which hypothesis is
true based on $n$ outputs $y_1, \ldots, y_n$.
In the following, randomized tests are allowed as our decision.
Hence, our decision method is described by a $[0,1]$-valued function $f$ on ${\cal Y}^n$.
When we observe $n$ outputs $y_1, \ldots, y_n$,
we accept the alternative hypothesis $\overline{P}$ with the probability $f(y_1, \ldots, y_n)$.
We have two types of errors.
In the first type, 
the null hypothesis $P$ is rejected despite being correct.
In the second type, 
the alternative $\overline{P}$ is rejected despite being correct.
Hence, the first type of error probability is given by ${\rm E}_{P^n}f$,
and the second type of error probability is by ${\rm E}_{{\overline{P}}^n}(1-f)$.
Note that ${\rm E}_{P}$ describes the expectation under the distribution $P$.

In the following, we assume that
\begin{align*}
\Phi(s|P\|\overline{P})
&:=
\int_{{\cal Y}} (\frac{\partial \overline{P}}{\partial P}(y))^s P(d y)<\infty \\
\phi(s|P\|\overline{P})
&:=\log \Phi(s|P\|\overline{P})
\end{align*}
and $\phi(s|P\|\overline{P})$ is $C^2$-continuous.
In the present paper, we choose the base of the logarithm to be $e$. 
In the discrimination of two distributions,
we treat two types of probabilities equally.
Then, we simply minimize the equal sum ${\rm E}_{P^n}f+{\rm E}_{{\overline{P}}^n}(1-f)$.
Its optimal rate of exponential decrease is
characterized by the Chernoff bound\cite{Chernoff}:
\begin{align*}
C(P,\overline{P}):=
\lim_{n\rightarrow\infty}
\frac{-1}{n}\log (\min_{f_n} {\rm E}_{P^n}f_n+{\rm E}_{{\overline{P}}^n}(1-f_n))
=-\min_{0 \le s \le 1}\phi(s|P\|\overline{P}).
\end{align*}
In order to treat these two error probabilities asymmetrically,
we often restrict the first type of error probability ${\rm E}_{P^n}f$ 
to below a particular threshold $\epsilon$,
and minimize the second type of error probability ${\rm E}_{{\overline{P}}^n}(1-f)$:
\begin{align*}
\beta_n^*(\epsilon):=\min_f
\bigl\{ {\rm E}_{{\overline{P}}^n}(1-f) &\bigm| 
{\rm E}_{P^n}f \le\epsilon \bigr\}.
\end{align*}
Then, the Stein's lemma holds.
For $0<\forall\epsilon<1$, the equation
\begin{align}
\lim_{n\rightarrow\infty}\frac{1}{n}\log\beta_n^*(\epsilon)
=-D(P\| \overline{P})
\Label{Stein}
\end{align}
holds, where the relative entropy $D(P\|\overline{P})$ is defined by
\begin{align*}
D(P\|\overline{P})=\int_{{\cal Y}} -\log 
\frac{\partial \overline{P}}{\partial P}(y) P(d y).
\end{align*}
Indeed, this lemma has the following variant form.
Define
\begin{align*}
B(P\|\overline{P})
:= &
\sup_{\{f_n\}}
\left\{ \left.
\varliminf_{n\rightarrow\infty}   \frac{-\log {\rm E}_{{\overline{P}}^n}(1-f_n)}{n} 
\right|
\lim_{n\rightarrow\infty} {\rm E}_{P^n} f_n =0
\right\}\\
B^*(P\|\overline{P})
:= &
\inf_{\{f_n\}}
\left\{ \left.
\varliminf_{n\rightarrow\infty}   \frac{-\log {\rm E}_{{\overline{P}}^n}(1-f_n)}{n} 
\right|
\varliminf_{n\rightarrow\infty}  {\rm E}_{P^n} f_n <1
\right\}.
\end{align*}
Then,
these two quantities satisfy the following relations:
\begin{align*}
B(P\|\overline{P})=B^*(P\|\overline{P})=D(P\|\overline{P}).
\end{align*}
As a further analysis,
we focus on the decreasing exponent of the error probability of the first type 
under an exponential constraint for the error probability of the second type.
When the decreasing exponent of for the error probability of the second type
is greater than the relative entropy $D(P\|\overline{P})$ ,
the error probability of the second type converges to $1$.
In this case, we focus on the decreasing exponent of the probability of correctly accepting the null hypothesis $P$.
For this purpose, we define 
\begin{align*}
B_e(r|P\|\overline{P})
:= &
\sup_{\{f_n\}}
\left\{ \left.
\varliminf_{n\rightarrow\infty}   \frac{-\log {\rm E}_{{P}^n} f_n }{n} 
\right|
\varliminf_{n\rightarrow\infty}   \frac{-\log {\rm E}_{{\overline{P}}^n}(1-f_n) }{n}\ge r
\right\}\\
B_e^*(r|P\|\overline{P})
:= &
\inf_{\{f_n\}}
\left\{ \left.
\varliminf_{n\rightarrow\infty}   \frac{-\log {\rm E}_{{P}^n}(1-f_n)}{n} 
\right|
\varliminf_{n\rightarrow\infty}   \frac{-\log {\rm E}_{{\overline{P}}^n}(1-f_n) }{n}\ge r
\right\}.
\end{align*}
Then, the two quantities are calculated as 
\begin{align}
B_e(r|P\|\overline{P})
&
=\min_{Q:D(Q\|\overline{P})\le r}D(Q\|P)
=\sup_{0\le s \le 1}\frac{-s r- \phi(s|P \|\overline{P})}{1-s}
\Label{1-18-1} \\
B_e^*(r|P\|\overline{P})
&=\min_{Q:D(Q\|\overline{P})\le r}D(Q\|P)+r-D(Q\|\overline{P})
=\sup_{s \le 0}\frac{-s r- \phi(s|P \|\overline{P})}{1-s}.\Label{22-1}
\end{align}
The first expressions of 
(\ref{1-18-1}) and (\ref{22-1}) are illustrated by Figs. \ref{f10} and \ref{f11}.
\begin{figure}[htbp]
\begin{center}
\scalebox{1.0}{\includegraphics[scale=0.4]{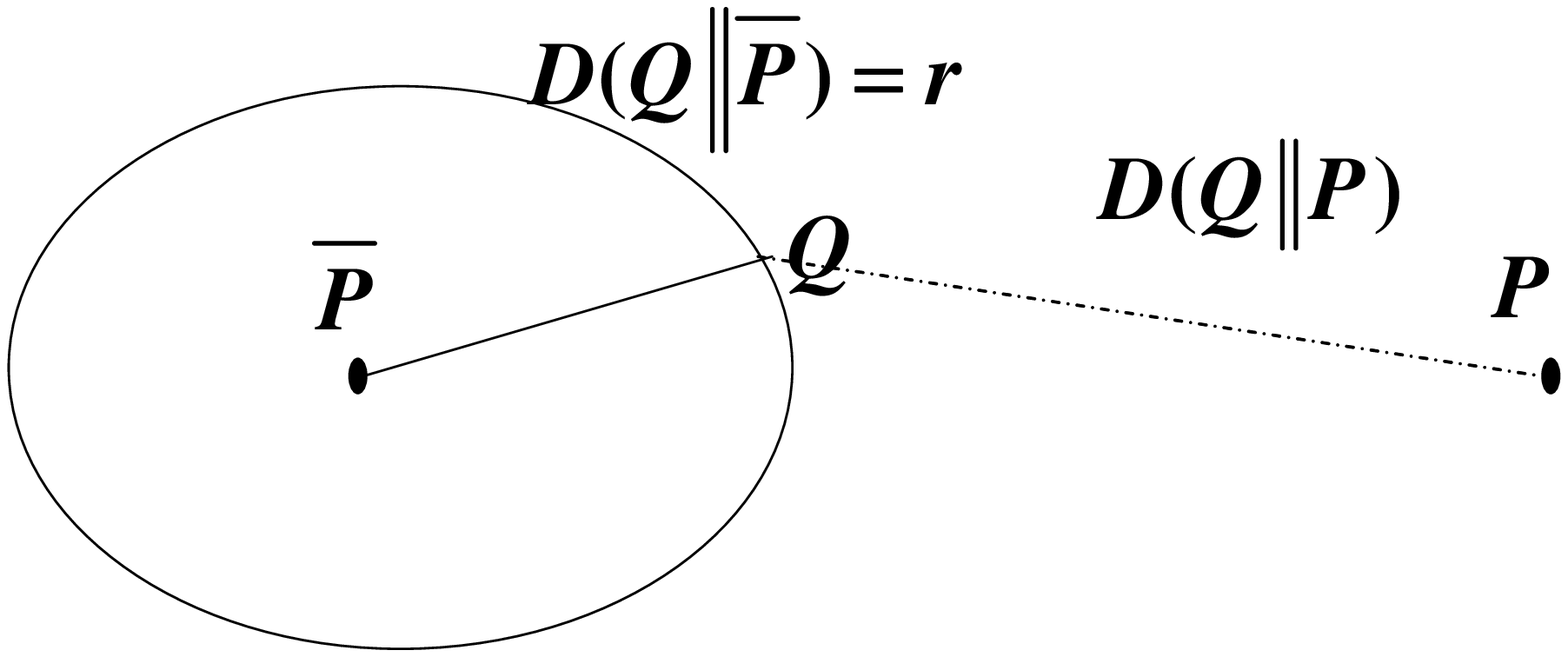}}
\end{center}
\caption{Figure of $B_e(r|P\|\overline{P})$}
\Label{f10}
\end{figure}%

\begin{figure}[htbp]
\begin{center}
\scalebox{1.0}{\includegraphics[scale=0.4]{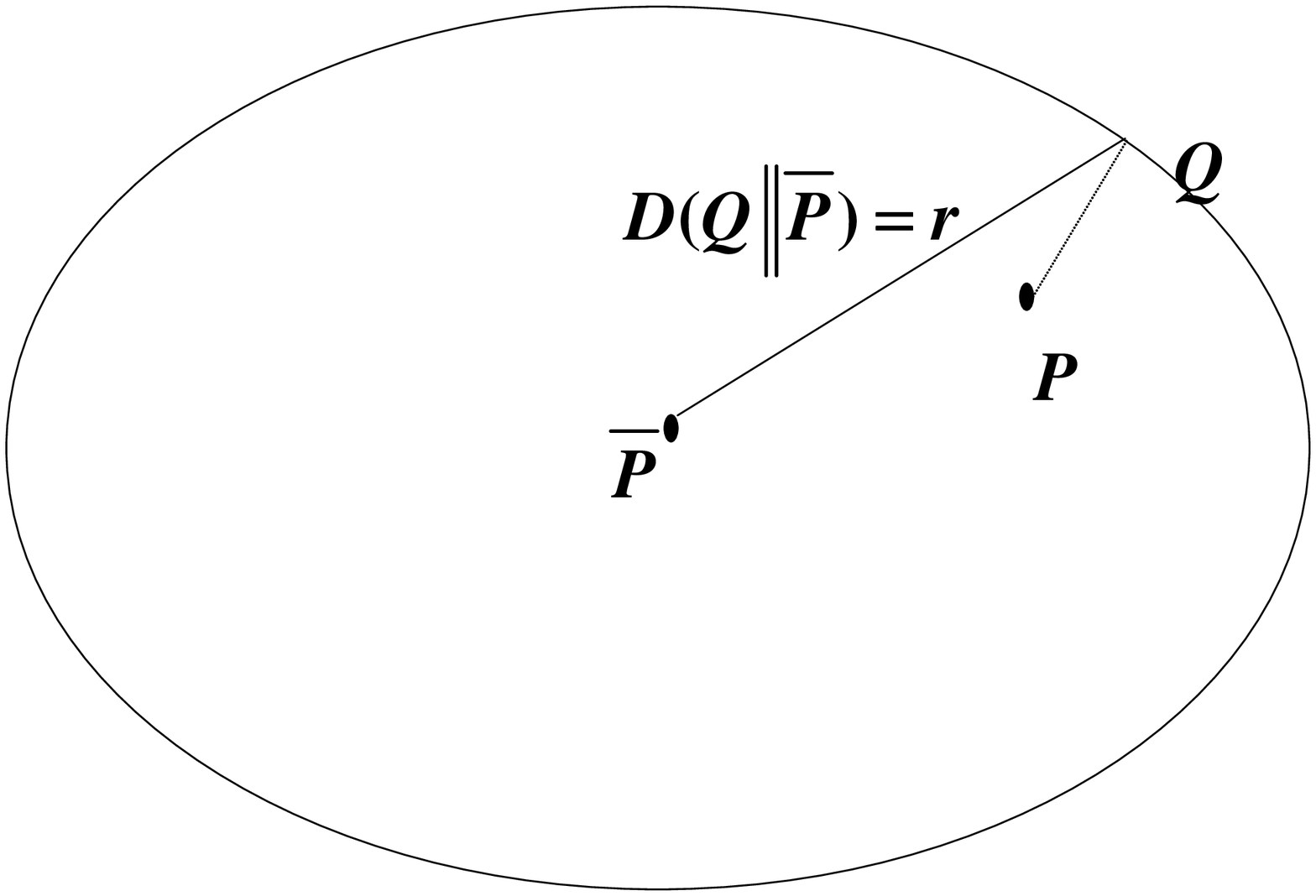}}
\end{center}
\caption{Figure of $B_e^*(r|P\|\overline{P})$ when $r0\ge r\ge D(P\|\overline{P})$}
\Label{f11}
\end{figure}%

Now, we define the new function $\overline{B}(r)$:
\begin{align*}
\overline{B}_e(r):= \left\{
\begin{array}{ll}
B_e(r|P\|\overline{P}) & r \le D(P\|\overline{P}) \\
- B_e^*(r|P\|\overline{P})& r > D(P\|\overline{P}).
\end{array}
\right.
\end{align*}
Then, its graph is shown in Fig. \ref{f3}.
\begin{figure}[htbp]
\begin{center}
\scalebox{1.0}{\includegraphics[scale=0.4]{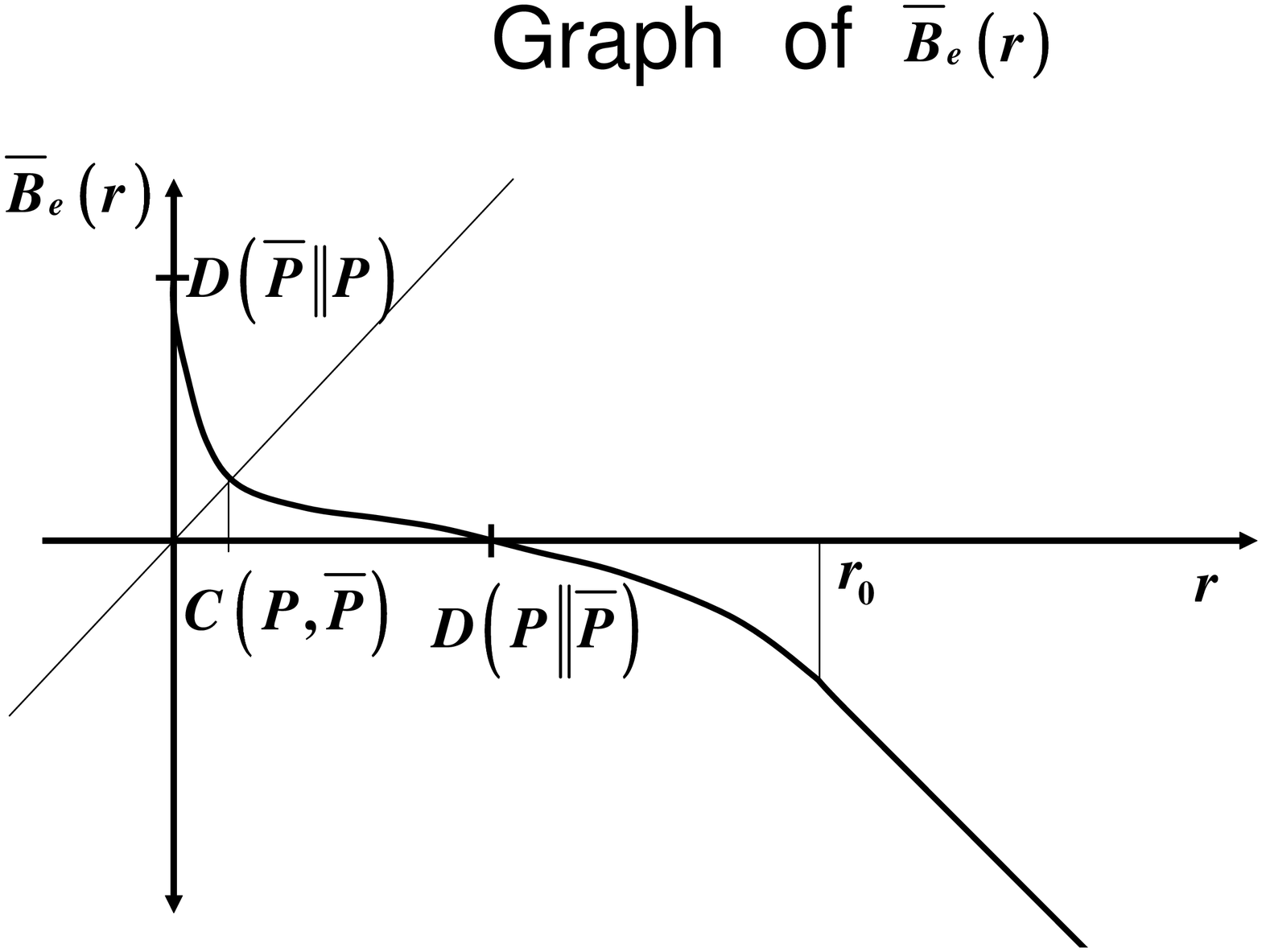}}
\end{center}
\caption{Graph of $B_e(r)$}
\Label{f3}
\end{figure}%

In order to give other characterizations of (\ref{1-18-1}),
we introduce a one-parameter family 
\begin{align*}
P_{s,P,\overline{P}}(dy):=
\frac{1}{\Phi(s|P\|P)}
(\frac{\partial \overline{P}}{\partial P}(y))^s P(d y),
\end{align*}
which is abbreviated as $P_{s}$. Then, since $\phi(s)$ is $C^1$ continuous,
\begin{align}
D(P_s\|P_1)&= (s-1)\phi'(s)-\phi(s) \quad s \in (-\infty,1]\Label{24-6} \\
D(P_0\|P_s)&= \phi(s)-s \phi'(0) \quad s \in [0,\infty)\Label{24-7}.
\end{align}
%D(P_s\|P_0)= -\phi(s)+s \phi'(s).
%That is, since $\phi(s)$ is $C^2$ continuous, there exists $\exists t(s)\in (0,s)$ or $(s,0)$ such that 
%\begin{align*}
%D(P_0\|P_s)= \phi(s)-s \phi'(0)= \frac{s^2}{2} \phi''(t(s)).
%\end{align*}
%In particular, 
%\begin{align}
%\phi''(t)= {\rm E}_{P_t}(\log \frac{\partial \overline{P}}{\partial P}(y))^2
%-({\rm E}_{P_t}\log \frac{\partial \overline{P}}{\partial P}(y))^2.
%\end{align}
Since
\begin{align*}
\frac{d (s-1)\phi'(s)-\phi(s)}{ds}=-\phi''(s) <0,
\end{align*}
$D(P_s\|P_1)$ is monotonically decreasing with respect to $s$.

As is mentioned in Theorem 4 of Blahut \cite{blahut}, 
when $r \le D(P\|\overline{P})$, 
there exists $s_r \in [0,1]$ such that
\begin{align*}
\min_{Q:D(Q\|\overline{P})\le r}D(Q\|P)=D(P_{s_r}\|P_0).
\end{align*}
Then, (\ref{24-6}) and (\ref{24-7}) imply that
\begin{align*}
r=D(P_{s_r}\|P_1)=(s_r-1)\phi(s_r)-\phi(s_r).
\end{align*}
Thus, we obtain another expression.
\begin{align}
\min_{Q:D(Q\|\overline{P})\le r}D(Q\|P)
= \min_{s \in [0,1]:D(P_{s}\|\overline{P})\le r}D(P_{s}\|P).
\Label{1-26-8} 
\end{align}
On the other hand, 
\begin{align}
\frac{d}{ds} 
\frac{-s r- \phi(s|P \|\overline{P})}{1-s}
=\frac{-r+ (s-1)\phi'(s)-\phi(s)}{(1-s)^2}
= \frac{D(P_s\|P_1)}{(1-s)^2}.\Label{24-2}
\end{align}
Since $D(P_s\|P_1)$ is monotonically decreasing with respect to $s$,
$\frac{d}{ds} 
\frac{-s r- \phi(s|P \|\overline{P})}{1-s}=0$ if and only if 
$s=s_r$.
The equation
\begin{align}
\min_{Q:D(Q\|\overline{P})\le r}D(Q\|P)
=\sup_{0\le s \le 1}\frac{-s r- \phi(s|P \|\overline{P})}{1-s}
\Label{24-1} 
\end{align}
can be checked.

In the following, we present some explanations concerning (\ref{22-1}).
As is mentioned by Han-Kobayashi\cite{HK} and Ogawa-Nagaoka\cite{Oga-Nag:test}, 
when $r_0 :=D(P_{-\infty}\|P_1) \ge r \ge D(P\|\overline{P})$, the relation
\begin{align*}
B_e^*(r|P\|\overline{P})=D(P_{s_r}\|P_0)
\end{align*}
holds, where $s_r\in (-\infty,0]$ is defined as
\begin{align*}
r=D(P_{s_r}\|P_1)=(s_r-1)\phi(s_r)-\phi(s_r).
\end{align*}
Thus, similar to (\ref{1-26-8}) and (\ref{24-1}),
the relation
\begin{align}
\min_{Q:D(Q\|\overline{P})\le r}D(Q\|P)+r-D(Q\|\overline{P})
&= 
D(P_{s_r}\|P)
=\sup_{s \le 0}\frac{-s r- \phi(s|P \|\overline{P})}{1-s}
\Label{1-26-8-1} 
\end{align}
holds, where $s_r\le 0$ is defined by $D(P_{s_r}\|\overline{P})= r$\cite{Oga-Nag:test}.

As mentioned by Nakagawa-Kanaya\cite{Nakagawa},  
when $r \ge r_0$, the relation
\begin{align*}
\min_{Q:D(Q\|\overline{P})\le r}D(Q\|P)+r-D(Q\|\overline{P})
= D(P_{-\infty}\|P)+r-D(P_{-\infty}\|\overline{P})
=
\min_{Q:D(Q\|\overline{P})\le r_0}(D(Q\|P)+r_0-D(Q\|\overline{P}))
+ r-r_0
\end{align*}
holds. 
This bound is attained by the following randomized test. The hypothesis $P$ is accepted with the probability only when the logarithmic likelihood ratio takes the maximum value $r_0$.
Since $D(P_s\|P_1)<r $, (\ref{24-2}) implies that
\begin{align}
& \sup_{s \le 0}\frac{-s r- \phi(s|P \|\overline{P})}{1-s}
=
\lim_{s \le -\infty}\frac{-s r- \phi(s|P \|\overline{P})}{1-s}
=
\lim_{s \le -\infty}\frac{-s r_0- \phi(s|P \|\overline{P})}{1-s}
+ r-r_0 \nonumber \\
= &
\min_{Q:D(Q\|\overline{P})\le r_0}(D(Q\|P)+r_0-D(Q\|\overline{P}))
+ r-r_0.
\end{align}
\Label{1-26-8-1} 

\begin{rem}
The classical Hoeffding bound in information
theory is due to Blahut\cite{blahut} and Csisz\'{a}r-Longo\cite{csiszarLongo}. The corresponding ideas in statistics
were first put forward by Hoeffding\cite{Hoe}, from whom the bound received
its name. Some authors prefer to refer this bound as the Hoeffding-Blahut-Csisz\'{a}r-
Longo bound.

On the other hand, 
Han-Kobayashi\cite{HK} gave the first equation of (\ref{22-1}),
and proved that this equation among non-randomized tests when $r_0 \ge r \ge D(P\|\overline{P})$.
They pointed out that the minimum 
$\min_{Q:D(Q\|\overline{P})\le r}D(Q\|P)+r-D(Q\|\overline{P})$
can be attained by $Q$ satisfying $D(Q\|\overline{P})= r$.
Ogawa-Nagaoka\cite{Oga-Nag:test}showed the second equation of (\ref{22-1})
for this case.

Nakagawa-Kanaya\cite{Nakagawa} proved the first equation when $r > r_0$.
Indeed, as pointed by Nakagawa-Kanaya\cite{Nakagawa}, 
when $r >  r_0$,
any non-randomized test cannot attain the minimum 
$\min_{Q:D(Q\|\overline{P})\le r}D(Q\|P)+r-D(Q\|\overline{P})$.
In this case, 
the minimum $\min_{Q:D(Q\|\overline{P})\le r}D(Q\|P)+r-D(Q\|\overline{P})$
cannot be attained by $Q$ satisfying $D(Q\|\overline{P})= r$.
\end{rem}

\section{Main result: Adaptive method}\Label{s3}
Let us focus on two spaces, the set of input signals ${\cal X}$ and the set of outputs ${\cal Y}$.
In this case, the channel from ${\cal X}$ and ${\cal Y}$ is described by 
the map from the set ${\cal X}$ to the set of probability distributions on ${\cal Y}$.
That is, given a channel $W$ 
$W_x$ represents the output distribution when the input is $x \in {\cal X}$.
When ${\cal X}$ and ${\cal Y}$ have finite elements,
the channel is given by transition matrix.
The main topic is the discrimination of two classical channels $W$ and $\overline{W}$.
In particular, we treat its asymptotic analysis when we can use the unknown channel only $n$ times.
That is, we discriminate two hypotheses,
the null hypothesis $H_0$ : $W^n$ versus the alternative hypothesis $H_1$: ${\overline{W}}^n$, 
where $W^n$ and ${\overline{W}}^n$ are the $n$ uses of the channel $W$ and $\overline{W}$
Then, our problem is to decide which hypothesis is
true based on 
$n$ inputs $x_1, \ldots, x_n$ and
$n$ outputs $y_1, \ldots, y_n$.
In this setting, it is allowed to choose the $k$-th input based on the previous $k-1$ output adaptively.
We choose the $k$-th input $x_k$ subject to the distribution 
$P^k_{(x_1,y_1), \ldots, (x_{k-1},y_{k-1})}(x_k)$ on ${\cal X}$.
That is, 
the $k$-th input $x_k$ 
depends on
$k$ conditional distributions $\vec{P}^k=(P^1, P^2, \ldots, P^k)$.
Hence, our decision method is described by 
$n$ conditional distributions $
\vec{P}^n=(P^1, P^2, \ldots, P^n)$ and 
a $[0,1]$-valued function $f_n$ on $({\cal X} \times {\cal Y})^n$.
In this case,
when we choose $n$ inputs $x_1, \ldots, x_n$ and 
observe $n$ outputs $y_1, \ldots, y_n$,
we accept the alternative hypothesis $\overline{W}$ with the probability 
$f_n(x_1, y_1, \ldots, x_n, y_n)$.
That is, our scheme is illustrated by Fig. \ref{f4}.
\begin{figure}[htbp]
\begin{center}
\scalebox{1.0}{\includegraphics[scale=0.4]{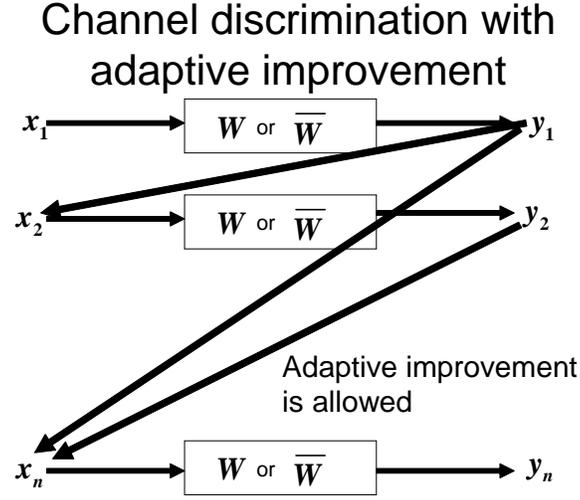}}
\end{center}
\caption{The adaptive method}
\Label{f4}
\end{figure}%

In order to treat this problem mathematically,
we introduce the following notation.
For a channel $W$ from ${\cal X}$ to ${\cal Y}$
and a distribution $P$ on ${\cal X}$,
we define two notations, the distribution $WP$ on ${\cal X}\times{\cal Y}$
and the distribution $W\cdot P$ on ${\cal Y}$ as 
\begin{align*}
W P(x,y)&:=W_x(y)P(x) \\
W\cdot P(x,y)&:=\int_{{\cal X}} W_x(y)P(dx).
\end{align*}
Using the distribution $WP$, we define two quantities:
\begin{align*}
D(W\|\overline{W}|P)
&:=D(W P\|\overline{W}P)\\
\phi(s|W\|\overline{W}|P)
&:=\phi(s|WP\|\overline{W}P).
\end{align*}
Based on $k$ conditional distributions $\vec{P}^k=(P^1, P^2, \ldots, P^k)$,
we define the following distributions:
\begin{align*}
Q_{W,\vec{P}^n}
&:=
W P^n W P^{n-1}\cdots W P^1\\
P_{W,\vec{P}^{n}}
&:=
P^n \cdot Q_{W,\vec{P}^{n-1}}\\
Q_{s,W|\overline{W},\vec{P}^n}
&:=P_{s,Q_{W,\vec{P}^n},Q_{\overline{W},\vec{P}^n}}\\
P_{s,W|\overline{W},\vec{P}^n}
&:= P^n \cdot Q_{s,W|\overline{W},\vec{P}^{n-1}}.
\end{align*}
Then, the first type of error probability is given by 
${\rm E}_{Q_{W,\vec{P}^n}}f_n$,
and the second type of error probability is by ${\rm E}_{Q_{\overline{W},\vec{P}^n}}(1-f_n)$.
In order to treat this problem,
we introduce the following quantities:
\begin{align*}
C(W,\overline{W})
&:=
\lim_{n\rightarrow\infty}
\frac{-1}{n}\log (\min_{\vec{P}^n, f_n} {\rm E}_{Q_{W,\vec{P}^n}}f_n
+{\rm E}_{Q_{\overline{W},\vec{P}^n}}(1-f_n))\\
\beta_n^*(\epsilon)
&:=\min_{\vec{P}^n, f_n}
\bigl\{ 
{\rm E}_{Q_{\overline{W},\vec{P}^n}}(1-f_n)
\bigm| 
{\rm E}_{Q_{W,\vec{P}^n}}f_n
\le\epsilon \bigr\},
\end{align*}
and
\begin{align*}
B(W\|\overline{W})
:= &
\sup_{\{(\vec{P}^n,f_n)\}}
\left\{ \left.
\varliminf_{n\rightarrow\infty}   \frac{-\log {\rm E}_{Q_{\overline{W},\vec{P}^n}}(1-f_n)}{n} 
\right|
\lim_{n\rightarrow\infty} {\rm E}_{Q_{W,\vec{P}^n}} f_n =0
\right\}\\
B^*(W\|\overline{W})
:= &
\inf_{\{(\vec{P}^n,f_n)\}}
\left\{ \left.
\varliminf_{n\rightarrow\infty}   \frac{-\log {\rm E}_{Q_{\overline{W},\vec{P}^n}}(1-f_n)}{n} 
\right|
\varliminf_{n\rightarrow\infty}  {\rm E}_{Q_{W,\vec{P}^n}} f_n <1
\right\}\\
B_e(r|W\|\overline{W})
:= &
\sup_{\{(\vec{P}^n,f_n)\}}
\left\{ \left.
\varliminf_{n\rightarrow\infty}   \frac{-\log {\rm E}_{Q_{W,\vec{P}^n}} f_n}{n} 
\right|
\varliminf_{n\rightarrow\infty}   \frac{-\log {\rm E}_{Q_{\overline{W},\vec{P}^n}} (1-f_n) }{n}\ge r
\right\}\\
B_e^*(r|W\|\overline{W})
:= &
\inf_{\{(\vec{P}^n,f_n)\}}
\left\{ \left.
\varliminf_{n\rightarrow\infty}   \frac{-\log {\rm E}_{Q_{W,\vec{P}^n}} (1-f_n)}{n} 
\right|
\varliminf_{n\rightarrow\infty}   \frac{-\log {\rm E}_{Q_{\overline{W},\vec{P}^n}} (1-f_n)}{n}\ge r
\right\}.
\end{align*}

We obtain the following channel version of Stein's lemma.
\begin{thm}\Label{Th1}
Assume that
$\phi(s|W_x \|\overline{W}_x)$ 
is $C^1$ continuous, and
\begin{align}
\lim_{\epsilon \to +0}
\frac{\phi(-\epsilon|W\|\overline{W})}{\epsilon}
=\sup_{x \in {\cal X}} D(W_x\|\overline{W}_x),\Label{2-10-1}
\end{align}
where $\phi(s|W\|\overline{W}):=
\sup_{x \in {\cal X}}\phi(s|W_x|\overline{W}_x)=\sup_{P \in {\cal P}({\cal X})} \phi(s|W\|\overline{W}|P)$, and ${\cal P}({\cal X})$ is the set of distributions on ${\cal X}$.

Then, 
\begin{align}
B(W\|\overline{W})
=B^*(W\|\overline{W})=\overline{D}:=\sup_{x \in {\cal X}}  D(W_x\|\overline{W}_x)
\Label{1-16-1}.
\end{align}
\end{thm}
The following is another expression of Stein's lemma.
\begin{cor}
Under the same assumption,
\begin{align*}
\lim_{n\rightarrow\infty}
\frac{-1}{n}\log\beta_n^*(\epsilon) 
&=\sup_{x \in {\cal X}}  D(W_x\|\overline{W}_x).
\end{align*}
\end{cor}

Condition (\ref{2-10-1}) can be replaced by another condition.
\begin{lem}\Label{2-10-2}
When any element $x \in {\cal X}$ satisfies
\begin{align*}
\phi'(0|W_x\|\overline{W}_x)
=D(W_x\|\overline{W}_x)
\end{align*}
and there exists a real number $\epsilon>0$ such that
\begin{align}
C_1:=\sup_{x \in {\cal X}}
\sup_{s\in [-\epsilon,0 ]}
\frac{d^2 \phi(s|W_x\|\overline{W}_x)}{ds^2}&< \infty \Label{1-26-4},
\end{align}
then condition (\ref{2-10-1}) holds.
\end{lem}

In addition, we obtain a channel version of the Hoeffding bound.
\begin{thm}\Label{Th2}
When 
\begin{align}
\sup_{x \in {\cal X}}
\sup_{s \in [0,1]}
\frac{d^2 \phi(s|W_x\|\overline{W}_x)}{ds^2}&< \infty \Label{1-26-4-a}
\end{align}
and 
\begin{align*}
\sup_{x \in {\cal X}}  D(\overline{W}_x\|W_x) < \infty,
\end{align*}
%Further, if the function 
%$r \mapsto
%\max_x \min_{Q:D(Q\|\overline{W}_x)\le r}D(Q\|W_x) $
%is uniform continuous concerning $x$ and , 
then
\begin{align}
B_e(r|W\|\overline{W})
&=
\sup_{x\in {\cal X}} \sup_{0\le s \le 1}\frac{-s r- \phi(s|W_x \|\overline{W}_x)}{1-s}
= \sup_{x\in {\cal X}} \min_{Q:D(Q\|\overline{W}_x)\le r}D(Q\|W_x) \Label{1-16-2}.
\end{align}
\end{thm}

\begin{cor}
Under the same assumption,
\begin{align}
C(W,\overline{W})
&=\sup_{x\in {\cal X}} -\min_{0 \le s\le 1} \phi(s|W_x \|\overline{W}_x)\Label{1-26-9}.
\end{align}
\end{cor}

These arguments imply that
adaptive improvement does not improve the performance in the above senses.
For example,
when we apply the best input $x_M:=\argmax_x D(W_x\|\overline{W}_x)$ to all of $n$ channels,
we can achieve the optimal performance in the sense of the Stein bound.
The same fact is true concerning the Hoeffding bound and the Chernoff bound.

\begin{proof}
The relation 
\begin{align*}
C(W,\overline{W})=
\sup \{r|B_e(r|W\|\overline{W})\ge r\}
\end{align*}
holds.
Since
\begin{align*}
& \sup \Bigl\{r\Bigl|
\sup_{x\in {\cal X}} \sup_{0\le s \le 1}\frac{-s r- \phi(s|W_x \|\overline{W}_x)}{1-s}
\ge r\Bigr.\Bigr\}\\
=&
\sup_{x\in {\cal X}} \sup \Bigl\{r\Bigl|
\sup_{0\le s \le 1}\frac{-s r- \phi(s|W_x \|\overline{W}_x)}{1-s}
\ge r\Bigr.\Bigr\}\\
=&
\sup_{x\in {\cal X}} 
-\min_{0 \le s\le 1} \phi(s|W_x \|\overline{W}_x),
\end{align*}
the relation (\ref{1-26-9}) holds.
\end{proof}

The channel version of the Han-Kobayashi bound is given as follows.
\begin{thm}\Label{Th3}
When 
$\phi(s|W_x \|\overline{W}_x)$ 
is $C^1$ continuous,
then
\begin{align}
B_e^*(r|W\|\overline{W})
&=\sup_{s \le 0}\frac{-s r- \phi(s|W \|\overline{W})}{1-s}
=
\inf_{P \in{\cal P}({\cal X})}
\sup_{s \le 0}\frac{-s r- \phi(s|W\|\overline{W}|P)}{1-s}
=
\inf_{P\in{\cal P}^2({\cal X}) }
\sup_{s \le 0}\frac{-s r- \phi(s|W\|\overline{W}|P)}{1-s},
\Label{1-16-3}
\end{align}
where 
${\cal P}^2({\cal X})$ 
is the distribution on ${\cal X}$ that takes positive probability only on at most two elements.
\end{thm}
As shown in Section \ref{s4},
the equality
\begin{align}
\sup_{s \le 0}\frac{-s r- \phi(s|W \|\overline{W})}{1-s}
=
\inf_{x \in {\cal X}}
\sup_{s \le 0}\frac{-s r- \phi(s|W_x\|\overline{W}_x)}{1-s} \Label{2-8-2}
\end{align}
does not necessarily hold in general.
In order to understand the meaning of this fact,
we assume that the equation (\ref{2-8-2}) does not hold.
When we apply the same input $x$ to all channels,
the best performance cannot be achieved.
However, the best performance can be achieved by the following method.
Assume that 
the best input distribution $\argmax_{P \in {\cal P}^2({\cal X})}
\sup_{s \le 0}\frac{-s r- \phi(s|W\|\overline{W}|P)}{1-s}$
has the support $\{x,x'\}$, and the probabilities $\lambda$ and $1-\lambda$.
Then, applying $x$ or $x'$ to all channels with the probability $\lambda$ and $1-\lambda$,
we can achieve the best performance in the sense of the Han-Kobayashi bound.
That is, the structure of optimal strategy of the Han-Kobayashi bound
is more complex than those of the above cases.

\section{Simple example}\Label{s4}
In this section, we treat a simple example that does not satisfy (\ref{2-8-2}).
For four given parameters $p,q, a>1,b>1$, we define the channels $W$ and $\overline{W}$:
\begin{align*}
W_0(0):=a q, \quad
W_0(1):=1-a q, \\
\overline{W}_0(0):=q, \quad
\overline{W}_0(1):=1-q, \\
W_1(0):=bq, \quad
W_1(1):=1-bq, \\
\overline{W}_1(0):=q, \quad
\overline{W}_1(1):=1-q.
\end{align*}
Then, we obtain
\begin{align*}
\lim_{s \to - \infty}
\frac{\phi(s|W_0\|\overline{W}_0)}{s}=a,\\
\lim_{s \to - \infty}
\frac{\phi(s|W_1\|\overline{W}_1)}{s}=b.
\end{align*}
In this case,
\begin{align*}
D(W_0\|\overline{W}_0)=& a p \log a + (1-ap)\log \frac{1-ap}{1-p} \\
D(W_1\|\overline{W}_1)=& b q \log b + (1-bq)\log \frac{1-bq}{1-q} .
\end{align*}
When $a>b$ and 
$D(W_0\|\overline{W}_0) <D(W_1\|\overline{W}_1)$,
the magnitude relation between 
$\phi(s|W_0\|\overline{W}_0)$ and $\phi(s|W_1\|\overline{W}_1)$
on $(-\infty,0)$ depends on $s\in (-\infty,0)$.
For example, the case of $a=100,b=1.5,p=0.0001,q=0.65$
is shown in Fig. \ref{f1}.
In this case, 
$B_e^*(r|W_0\|\overline{W}_0)$, $B_e^*(r|W_1\|\overline{W}_1)$,
and 
$B_e^*(r|W\|\overline{W})$
are calculated by Fig. \ref{f2}.
Then, the inequality (\ref{2-8-2}) does not hold.

\begin{figure}[htbp]
\begin{center}
\scalebox{1.0}{\includegraphics[scale=1]{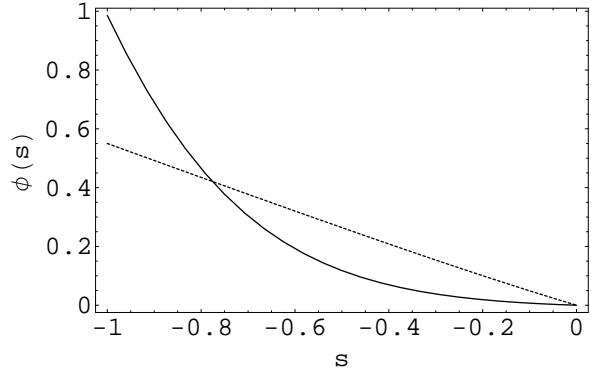}}
\end{center}
\caption{Magnitude relation between $\phi(s|W_0\|\overline{W}_0)$
and $\phi(s|W_1\|\overline{W}_1)$ on $(-1,0)$.
The upper solid line indicates $\phi(s|W_0\|\overline{W}_0)$, 
the dotted line indicates $\phi(s|W_1\|\overline{W}_1)$.}
\Label{f1}
\end{figure}%

\begin{figure}[htbp]
\begin{center}
\scalebox{1.0}{\includegraphics[scale=1]{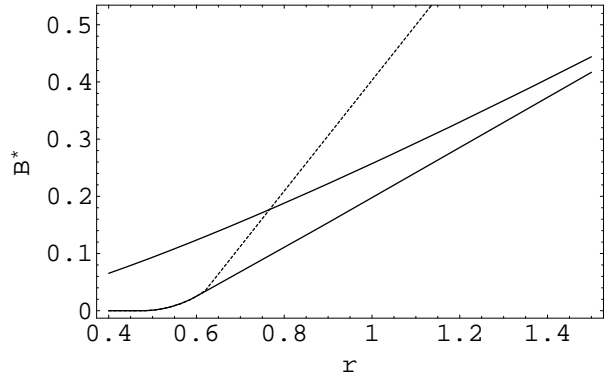}}
\end{center}
\caption{
Magnitude relation between $B_e^*(r|W_0\|\overline{W}_0)$, $B_e^*(r|W_1\|\overline{W}_1)$, and $B_e^*(r|W\|\overline{W})$ on $(-1,0)$.
The upper solid line indicates $B_e^*(r|W_0\|\overline{W}_0)$, 
the dotted line indicates $B_e^*(r|W_1\|\overline{W}_1)$, and
the lower solid line indicates $B_e^*(r|W\|\overline{W})$.}
\Label{f2}
\end{figure}%

\section{Application to adaptive quantum state discrimination}\Label{s5}
Quantum state discrimination between two states $\rho$ and $\sigma$ on a $d$-dimensional system ${\cal H}$
with $n$ copies by one-way LOCC is formulated as follows.
We choose the first POVM $M_1$ and obtain 
the data $y_1$ through the measurement $M_1$.
In the $k$-th step, we choose the $k$-th POVM $M_k((M_1,y_1),\ldots,(M_{k-1},y_{k-1}) )$ depending on $(M_1,y_1),\ldots,(M_{k-1},y_{k-1})$.
Then, we obtain the $k$-th data $y_k$ through $M_k((M_1,y_1),\ldots,(M_{k-1},y_{k-1}) )$.
Therefore, this problem can be regarded as classical channel discrimination with the correspondence 
$W_M(y)=\Tr M(y)\rho$ and 
$\overline{W}_M(y)=\Tr M(y)\sigma$.
That is, in this case, the set of input signal corresponds to the set of extremal points of the set of POVMs on the given system ${\cal H}$.
The proposed scheme is illustrated in Fig. \ref{f5}.
\begin{figure}[htbp]
\begin{center}
\scalebox{1.0}{\includegraphics[scale=0.4]{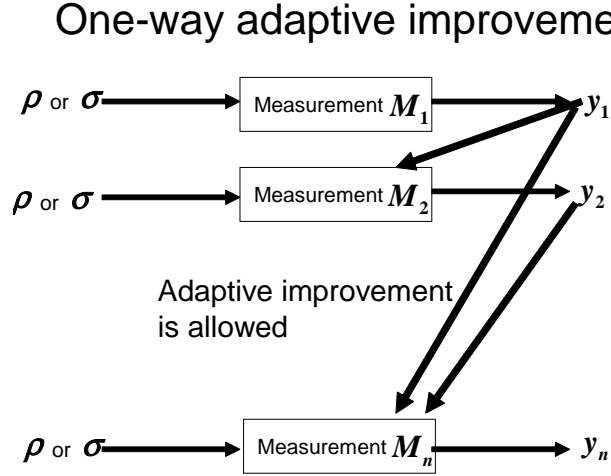}}
\end{center}
\caption{Adaptive quantum state discrimination}
\Label{f5}
\end{figure}%

Now, we assume that $\rho>0$ and $\sigma >0$.
In this case, 
${\cal X}$ is compact,
and 
the map $(s,M) \to \frac{d^2 \phi(s|W_M\|\overline{W}_M)}{ds^2}$
is continuous. Then, the condition (\ref{1-26-4}) holds.
Therefore, one-way improvement does not improve the performance in the sense of 
the Stein bound, the Chernoff bound, the Hoeffding bound, or the Han-Kobayashi bound.
That is, we obtain
\begin{align*}
B(W\|\overline{W})
=&
B^*(W\|\overline{W})
=\max_{M:{\rm POVM}}D(P^M_\rho\|P^M_\sigma)\\
B_e(r|W\|\overline{W})
=&
\max_{M:{\rm POVM}}
\sup_{0\le s \le 1}\frac{-s r- \phi(s|P^M_\rho\|P^M_\sigma)}{1-s} \\
B_e^*(r|W\|\overline{W})
=&
\sup_{s \le 0}\frac{-s r- \max_{M:{\rm POVM}}
\phi(s|P^M_\rho\|P^M_\sigma)}{1-s}.
\end{align*}
Therefore, there exists a difference between one-way LOCC and collective measurement.

\section{Proof of the Stein bound: (\ref{1-16-1})}\Label{s6}
Now, we prove the Stein bound: (\ref{1-16-1}).
For any $x \in {\cal X}$,
by choosing the input $x$ in $n$ times,
we obtain
\begin{align*}
B(W\|\overline{W})
\ge
D(W_x\|\overline{W}_x).
\end{align*}
Taking the supremum,
we have
\begin{align*}
B(W\|\overline{W})
\ge
\sup_{x \in {\cal X}} D(W_x\|\overline{W}_x).
\end{align*}
Furthermore, from the definition, it is trivial that
\begin{align*}
B(W\|\overline{W})\le B^*(W\|\overline{W}).
\end{align*}

Therefore, it is sufficient to show 
the strong converse part:
\begin{align}
B^*(W\|\overline{W})
\le
\overline{D}.
%(:=\sup_{x \in {\cal X}} D(W_x\|\overline{W}_x)).
\Label{1-25-2}
\end{align}
However, in preparation for the proof of (\ref{1-16-2}),
we present a proof of 
the weak converse part:
\begin{align}
B(W\|\overline{W})
\le
\overline{D} \Label{1-25-3}
\end{align}
which is weaker argument than (\ref{1-25-2}), and is valid without assumption (\ref{2-10-1}).
In the following proof, it is essential to evaluate 
the KL-divergence concerning the obtained data.

In order to prove (\ref{1-25-3}),
we prove that 
\begin{align}
\vlimsup_{n\to \infty}
-\frac{1}{n} \log {\rm E}_{Q_{\overline{W},\vec{P}^n}} (1-f_n) 
\le
\overline{D} \Label{1-25-4}
\end{align}
when 
\begin{align}
{\rm E}_{Q_{W,\vec{P}^n}} f_n \to 0.\Label{1-25-5}
\end{align}
It follows from the definitions of 
$Q_{W,\vec{P}^n}$ and 
$Q_{\overline{W},\vec{P}^n}$ that
\begin{align*}
D(Q_{W,\vec{P}^n}\|Q_{\overline{W},\vec{P}^n})
=\sum_{k=1}^n D(W\|\overline{W}|P_{W,\vec{P}^k}).
\end{align*}
Since $-{\rm E}_{Q_{W,\vec{P}^n}} f_n
\log
{\rm E}_{Q_{\overline{W},\vec{P}^n}} f_n \ge 0$,
information processing inequality concerning the KL divergence yields the following:
\begin{align}
& -h({\rm E}_{Q_{W,\vec{P}^n}} (1-f_n))
-({\rm E}_{Q_{W,\vec{P}^n}} (1-f_n))\log
{\rm E}_{Q_{\overline{W},\vec{P}^n}} (1-f_n) \nonumber \\
\le &
{\rm E}_{Q_{W,\vec{P}^n}} (1-f_n)
(\log 
{\rm E}_{Q_{W,\vec{P}^n}} (1-f_n)
-
\log 
{\rm E}_{Q_{\overline{W},\vec{P}^n}} (1-f_n))
+
{\rm E}_{Q_{W,\vec{P}^n}} f_n
(\log
{\rm E}_{Q_{W,\vec{P}^n}} f_n
-
\log
{\rm E}_{Q_{\overline{W},\vec{P}^n}} f_n
) \nonumber  \\
\le & D(Q_{W,\vec{P}^n}\|Q_{\overline{W},\vec{P}^n})
=\sum_{k=1}^n D(W\|\overline{W}|P_{W,\vec{P}^k})
\le n \overline{D}. \Label{1-25-9}
\end{align}
That is,
\begin{align}
-\frac{1}{n} \log {\rm E}_{Q_{\overline{W},\vec{P}^n}} (1-f_n) 
\le
\frac{
\overline{D}+\frac{1}{n}h({\rm E}_{Q_{W,\vec{P}^n}} (1-f_n))
}{{\rm E}_{Q_{W,\vec{P}^n}} (1-f_n)}.\Label{1-25-8}
\end{align}
Therefore, (\ref{1-25-5}) yields (\ref{1-25-4}).

Next, we prove the strong converse part, i.e., we show that
\begin{align}
{\rm E}_{Q_{W,\vec{P}^n}} (1-f_n) \to 0
\Label{1-25-1-a}
\end{align}
when
\begin{align}
& r:=\varliminf_{n\rightarrow\infty} 
 \frac{-\log {\rm E}_{Q_{\overline{W},\vec{P}^n}} (1-f_n)}{n}
>  \overline{D}.
\Label{1-25-1}
\end{align}
Since
\begin{align*}
&
\Phi(s|Q_{W,\vec{P}^n}\|Q_{\overline{W},\vec{P}^n})\\
=&
\Phi(s|Q_{W,\vec{P}^{n-1}}\|Q_{\overline{W},\vec{P}^{n-1}})
\left(
\int_{{\cal X}}
\left(\int_{{\cal Y}}
(\frac{\partial W_{x_n}'}{\partial W_{x_n}}(y_n))^s
W_{x_n}(d y_n)\right)
P_{s,W|\overline{W},\vec{P}^n}(d x_n)
\right),
\end{align*}
we obtain
\begin{align}
\phi(s|Q_{W,\vec{P}^n}\|Q_{\overline{W},\vec{P}^n})
=
\phi(s|Q_{W,\vec{P}^{n-1}}\|Q_{\overline{W},\vec{P}^{n-1}})
+\phi(s|W\|\overline{W}|P_{s,W|\overline{W},\vec{P}^n})\Label{1-16-6}.
\end{align}
Applying (\ref{1-16-6}) inductively,
we obtain the relation
\begin{align}
\phi(s|Q_{W,\vec{P}^n}\|Q_{\overline{W},\vec{P}^n})
=\sum_{k=1}^n \phi(s|W\|\overline{W}|P_{s,W|\overline{W},\vec{P}^k})
\le n \phi(s|W\|\overline{W}). \Label{25-2}
\end{align}
Since the information quantity $\phi(s|P \|\overline{P})$ satisfies the information processing inequality,
we have
\begin{align*}
&({\rm E}_{Q_{W,\vec{P}^n}} (1-f_n))^{1-s}
({\rm E}_{Q_{\overline{W},\vec{P}^n}} (1-f_n))^s \\
\le &
({\rm E}_{Q_{W,\vec{P}^n}} (1-f_n))^{1-s}
({\rm E}_{Q_{\overline{W},\vec{P}^n}} (1-f_n))^s
+
({\rm E}_{Q_{W,\vec{P}^n}} f_n)^{1-s}
({\rm E}_{Q_{\overline{W},\vec{P}^n}} f_n)^s\\
\le &
e^{\phi(s|Q_{W,\vec{P}^n}\|Q_{\overline{W},\vec{P}^n})}\\
\le &
e^{n \phi(s|W\|\overline{W})},
\end{align*}
for $s \le 0$.
Taking the logarithm, we obtain 
\begin{align}
(1-s) \log {\rm E}_{Q_{W,\vec{P}^n}} (1-f_n)
\le -s \log {\rm E}_{Q_{\overline{W},\vec{P}^n}} (1-f_n)
+ n \phi(s|W\|\overline{W}).\Label{2-5-2}
\end{align}
That is,
\begin{align*}
\frac{-1}{n}\log {\rm E}_{Q_{W,\vec{P}^n}} (1-f_n)
\ge 
\frac{-s \frac{-1}{n}\log {\rm E}_{Q_{\overline{W},\vec{P}^n}} (1-f_n)
- \phi(s|W\|\overline{W})}{1-s}.
\end{align*}
When $\varliminf_{n\rightarrow\infty}\frac{-\log {\rm E}_{{\overline{P}}^n}(1-f_n) }{n}\ge r$,
the inequality
\begin{align*}
B_e^*(r|W\|\overline{W})
\ge 
\varliminf_{n\rightarrow\infty} \frac{-1}{n}\log {\rm E}_{Q_{W,\vec{P}^n}} (1-f_n)
\ge 
\frac{-s r- \phi(s|W\|\overline{W})}{1-s}
\end{align*}
holds.
Taking the supremum, we obtain
\begin{align*}
B_e^*(r|W\|\overline{W})
\ge 
\sup_{s \le 0}\frac{-s r- \phi(s|W \|\overline{W})}{1-s}.
\end{align*}
From conditions (\ref{2-10-1}) and (\ref{1-25-1}),
there exists a small real number $\epsilon >0$ such that
$r > \frac{\phi(-\epsilon|W \|\overline{W})}{-\epsilon}$.
Thus, 
\begin{align*}
\sup_{s \le 0}\frac{-s r- \phi(s|W \|\overline{W})}{1-s}
\ge
\frac{\epsilon r- \phi(-\epsilon|W \|\overline{W})}{1+\epsilon} >0.
\end{align*}
Therefore, we obtain (\ref{1-25-1-a}).

\begin{rem}
The technique of the strong converse part except for (\ref{25-2}) was developed by 
Nagaoka \cite{Naga-EQIS}.
Hence, deriving (\ref{25-2}) can be regarded as the main contribution in this section of the present paper.
\end{rem}

Proof of Lemma \ref{2-10-2}:

It is sufficient for a proof of (\ref{2-10-1}) to show that
the uniformity of the convergence 
$\frac{\phi(-\epsilon|W_x\|\overline{W}_x)}{\epsilon}
-D(W_x\|\overline{W}_x)\to 0$
concerning $x \in {\cal X}$.
Now, we choose $\epsilon>0$ satisfying condition (\ref{1-26-4}).
Then, there exists $s \in [-\epsilon,0]$ such that
$\frac{\phi(-\epsilon|W_x\|\overline{W}_x)}{\epsilon}
-D(W_x\|\overline{W}_x)= \frac{1}{2}\epsilon \phi(s|W_x\|\overline{W}_x)\le \frac{C_1}{2}\epsilon$.
Therefore, 
the condition (\ref{2-10-1}) holds.

\section{Proof of the Hoeffding bound: (\ref{1-16-2})}\Label{s7}
In this section, we prove the Hoeffding bound: (\ref{1-16-2}).
Since the inequality
\begin{align*}
B_e(r|W\|\overline{W})
\ge
\sup_{x\in {\cal X}} \sup_{0\le s \le 1}\frac{-s r- \phi(s|W_x \|\overline{W}_x)}{1-s}
= \sup_{x\in {\cal X}} \min_{Q:D(Q\|\overline{W}_x)\le r}D(Q\|W_x) 
\end{align*}
is trivial,
we prove the opposite inequality.
In the following proof, the geometric characterization Fig. \ref{f10} and
the weak and the strong converse parts are essential.
Equation (\ref{1-26-8}) guarantees that
\begin{align*}
\sup_{x\in {\cal X}}  \min_{Q:D(Q\|\overline{W}_x)\le r}D(Q\|W_x) 
= \sup_{x\in {\cal X}} \min_{s \in [0,1]:D(P_{s,W_x,\overline{W}_x}\|\overline{W}_x)\le r}
D(P_{s,W_x,\overline{W}_x}\|W_x) .
\end{align*}
For this purpose, for arbitrary $\epsilon >0$,
we choose a channel $V:V_x=P_{s(x),W_x,\overline{W}_x}$ by
\begin{align*}
s(x)
:= \argmin_{s \in [0,1]:D(P_{s,W_x,\overline{W}_x}\|\overline{W}_x)\le r}
D(P_{s,W_x,\overline{W}_x}\|W_x) .
\end{align*}
Assume that a sequence $\{(\vec{P}^n,f_n)\}$ satisfies
\begin{align*}
\vlimsup_{n\to \infty}  \frac{-1}{n}\log {\rm E}_{Q_{\overline{W},\vec{P}^n}} (1-f_n)=r.
\end{align*}
By substituting $V$ into $W$,
the strong converse part of the Stein bound:(\ref{1-25-1-a}) implies that
\begin{align*}
\lim  {\rm E}_{Q_{V,\vec{P}^n}} (1-f_n)=0.
\end{align*}
The condition (\ref{1-26-4}) can be checked 
by the following relations:
\begin{align}
\frac{d \phi(t|P_{s(x),W_x,\overline{W}_x}\|\overline{W}_x)}{dt}
&=
(1-s(x))\phi'(s(x)(1-t)+t|W_x\|\overline{W}_x) \Label{1-26-5}\\
\frac{d^2 \phi(t|P_{s(x),W_x,\overline{W}_x}\|\overline{W}_x)}{dt^2}
&=
(1-s(x))^2\phi''(s(x)(1-t)+t|W_x\|\overline{W}_x).\Label{1-26-6}
\end{align}

Thus, by substituting $V$ and $W$ into $W$ and $\overline{W}$,  the relation (\ref{1-25-8}) implies that
\begin{align*}
\vlimsup_{n\to \infty}
-\frac{1}{n} \log {\rm E}_{Q_{\overline{W},\vec{P}^n}} (1-f_n) 
\le
\sup_{x \in {\cal X}}
D(V_x\|W_x) .
\end{align*}
Similar to (\ref{1-26-5}) and (\ref{1-26-6}), 
we can check the condition (\ref{1-26-4}).

From the construction of $V$,
we obtain
\begin{align*}
\vlimsup_{n\to \infty}
-\frac{1}{n} \log {\rm E}_{Q_{\overline{W},\vec{P}^n}} (1-f_n) 
\le
\max_x \min_{Q:D(Q\|\overline{W}_x)\le r-\epsilon}D(Q\|W_x) .
\end{align*}
The uniform continuity guarantees that
\begin{align*}
\vlimsup_{n\to \infty}
-\frac{1}{n} \log {\rm E}_{Q_{\overline{W},\vec{P}^n}} (1-f_n) 
\le
\max_x \min_{Q:D(Q\|\overline{W}_x)\le r}D(Q\|W_x) .
\end{align*}

Now, we show the uniformity of the function $r\mapsto 
\sup_{0\le s \le 1}\frac{-s r- \phi(s|W_x \|\overline{W}_x)}{1-s}
$ concerning $x$.
As mentioned in p. 82 of Hayashi\cite{Hayashi},
the relation
\begin{align*}
\frac{d}{dr}
\sup_{0\le s \le 1}\frac{-s r- \phi(s|W_x \|\overline{W}_x)}{1-s}=
\frac{s_r}{s_r-1}
\end{align*}
holds, where
\begin{align*}
s_r:=\argmax_{0\le s \le 1}\frac{-s r- \phi(s|W_x \|\overline{W}_x)}{1-s}.
\end{align*}
Since
\begin{align*}
\frac{d}{dr}\left.
\frac{-s r- \phi(s|W_x \|\overline{W}_x)}{1-s}\right|_{s=s_r}=0,
\end{align*}
we have
\begin{align*}
r=(s_r-1)\phi'(s_r|W_x \|\overline{W}_x)-\phi(s_r|W_x \|\overline{W}_x).
\end{align*}
Since $-\phi(s_r|W_x \|\overline{W}_x)\ge 0$, $(s_r-1) \le 0$, and
$\phi''(s|W_x \|\overline{W}_x)\ge 0$,
\begin{align*}
r\ge 
(s_r-1)\phi'(s_r|W_x \|\overline{W}_x)
\ge (s_r-1) \phi'(1|W_x \|\overline{W}_x)
= (1-s_r) D(\overline{W}_x\|W_x).
\end{align*}
Thus,
\begin{align*}
\frac{r}{D(\overline{W}_x\|W_x)} \ge (1-s_r).
\end{align*}
Hence,
\begin{align*}
|\frac{s_r}{s_r-1}|
\le \frac{1}{1-s_r}\le
\frac{D(\overline{W}_x\|W_x)}{r}
\le \frac{\sup_{x}D(\overline{W}_x\|W_x)}{r}.
\end{align*}
Therefore, 
the function $r\mapsto 
\sup_{0\le s \le 1}\frac{-s r- \phi(s|W_x \|\overline{W}_x)}{1-s}
$ is uniform continuous with respect to $x$.

\section{Proof of the Han-Kobayashi bound: (\ref{1-16-3})}\Label{s8}
The inequality
\begin{align}
B_e(r|W\|\overline{W})
\ge
\sup_{s \le 0}\frac{-s r- \phi(s|W\|\overline{W})}{1-s}.\Label{1-5-4}
\end{align}
has been shown in Section \ref{s6}, and the inequality
\begin{align*}
B_e(r|W\|\overline{W})
\le
\inf_{P\in {\cal P}^2({\cal X})}
\sup_{s \le 0}\frac{-s r- \phi(s|W\|\overline{W}|P)}{1-s}
\end{align*}
can be easily check by considering the input $P$.
Therefore, it is sufficient to show the inequality
\begin{align}
\inf_{P\in {\cal P}^2({\cal X})}
\sup_{s \le 0}\frac{-s r- \phi(s|W\|\overline{W}|P)}{1-s}
\le 
\sup_{s \le 0}
\frac{-s r- \phi(s|W\|\overline{W})}{1-s}
=
\sup_{s \le 0}
\inf_{P\in {\cal P}^2({\cal X})}
\frac{-s r- \phi(s|W\|\overline{W}|P)}{1-s}.
\Label{16-7}
\end{align}
This relation seems to be guaranteed by the mini-max theorem (Chap. VI Prop. 2.3 of \cite{Eke-Tem}).
However, the function $\frac{-s r- \phi(s|W\|\overline{W}|P)}{1-s}$ is not necessarily concave
concerning $s$ while it is convex concerning $P$.
Hence, this relation cannot be guaranteed by the mini-max theorem.

Now, we prove this inequality when the maximum $\max_{s \le 0}
\frac{-s r- \phi(s|W\|\overline{W})}{1-s}$ exists.
Since $\phi(s|W_x\|\overline{W}_x)$ is convex concerning $s$,
$\phi(s|W\|\overline{W})$ is also convex concerning $s$.
Then, we can define 
\begin{align*}
\partial^+ \phi(s|W\|\overline{W}):=
\lim_{\epsilon \to +0}\frac{\phi(s+\epsilon|W\|\overline{W})-\phi(s|W\|\overline{W})}{\epsilon} \\
\partial^- \phi(s|W\|\overline{W}):=
\lim_{\epsilon \to +0}\frac{\phi(s|W\|\overline{W})-\phi(s-\epsilon|W\|\overline{W})}{\epsilon}.
\end{align*}
Hence, the real number $s_r:=\argmax_{s \le 0}
\frac{-s r- \phi(s|W\|\overline{W})}{1-s}$ satisfies that
\begin{align*}
(1-s_r)\partial^- \phi(s_r|W\|\overline{W})+\phi(s_r|W\|\overline{W})
\le -r 
\le
(1-s_r)\partial^+ \phi(s_r|W\|\overline{W})+\phi(s_r|W\|\overline{W}).
\end{align*}
That is, there exists $\lambda \in [0,1]$ such that
\begin{align}
-r=(1-s_r)
(\lambda \partial^+ \phi(s_r|W\|\overline{W})+(1-\lambda)\partial^- \phi(s_r|W\|\overline{W}))
+\phi(s_r|W\|\overline{W}).\Label{19-3}
\end{align}
For an arbitrary real number $1> \epsilon >0$,
there exists $1> \delta>0$ such that 
\begin{align}
\frac{\phi(s+\delta|W\|\overline{W})-\phi(s|W\|\overline{W})}{\delta} 
& \le \partial^+ \phi(s|W\|\overline{W})+ \epsilon \Label{16-1} \\
\frac{\phi(s|W\|\overline{W})-\phi(s-\delta|W\|\overline{W})}{\delta}
& \ge \partial^- \phi(s|W\|\overline{W})- \epsilon.\Label{16-2}
\end{align}
Then, we choose $x^+, x^- \in {\cal X}$ such that
\begin{align}
\phi(s_r+\lambda \delta|W\|\overline{W}) - \delta \epsilon
\le
\phi(s_r+\lambda \delta|W_{x^+}\|\overline{W}_{x^+}) 
\le
\phi(s_r+\lambda \delta|W\|\overline{W}) \Label{16-3}
\\
\phi(s_r-(1-\lambda) \delta|W\|\overline{W}) - \delta \epsilon
\le
\phi(s_r-(1-\lambda) \delta|W_{x^-}\|\overline{W}_{x^-}) 
\le \phi(s_r-(1-\lambda) \delta|W\|\overline{W}).\Label{16-4}
\end{align}
Thus, (\ref{16-3}) implies that
\begin{align}
& \frac{
\phi(s_r+\lambda \delta|W_{x^+}\|\overline{W}_{x^+})
-
\phi(s_r-(1-\lambda) \delta|W_{x^+}\|\overline{W}_{x^+})
}{\delta}\nonumber \\
\ge &
\frac{
\phi(s_r+\lambda \delta|W\|\overline{W})-\delta \epsilon
-
\phi(s_r-(1-\lambda) \delta|W\|\overline{W})
}{\delta}\nonumber \\
\ge &
\frac{
\phi(s_r+\lambda \delta|W\|\overline{W})
-\phi(s_r+|W\|\overline{W})
+\phi(s_r+|W\|\overline{W})
-\phi(s_r-(1-\lambda) \delta|W\|\overline{W})
-\delta \epsilon
}{\delta}\nonumber \\
\ge &
\frac{
\lambda \delta \partial^+ \phi(s_r|W\|\overline{W})
+(1-\lambda) \delta (\partial^- \phi(s_r+|W\|\overline{W}) -\epsilon)
- \delta \epsilon
}{\delta}\nonumber \\
= &
\lambda \partial^+ \phi(s_r|W\|\overline{W})
+(1-\lambda) \partial^- \phi(s_r+|W\|\overline{W})
-\epsilon.\Label{19-8}
\end{align}
Similarly, (\ref{16-4}) implies that
\begin{align}
& \frac{
\phi(s_r+\lambda \delta|W_{x^-}\|\overline{W}_{x^-})
-
\phi(s_r-(1-\lambda) \delta|W_{x^-}\|\overline{W}_{x^-})
}{\delta}\nonumber \\
\le &
\lambda \partial^+ \phi(s_r|W\|\overline{W})
+(1-\lambda) \partial^- \phi(s_r+|W\|\overline{W})
+\epsilon. \Label{19-9}
\end{align}
Therefore, there exists a real number $\lambda'\in [0,1]$ such that
\begin{align}
& 
\left|
\frac{
\varphi(s_r+\lambda \delta|\lambda')
-
\varphi(s_r-(1-\lambda) \delta|\lambda')
}{\delta}
-(\lambda \partial^+ \phi(s_r|W\|\overline{W})
+(1-\lambda) \partial^- \phi(s_r+|W\|\overline{W}))\right|
\nonumber  \\
\le &
\epsilon.\Label{16-5}
\end{align}
where
\begin{align*}
\varphi(s|\lambda'):=
\lambda'\phi(s|W_{x^+}\|\overline{W}_{x^+})
+
(1-\lambda')\phi(s|W_{x^-}\|\overline{W}_{x^-}).
\end{align*}
Thus, there exists $\overline{s}_r\in [s_r-(1-\lambda)\delta,s_r+\lambda \delta]$
such that 
\begin{align}
\left|\varphi'(\overline{s}_r|\lambda')
-(\lambda \partial^+ \phi(s_r|W\|\overline{W})
+(1-\lambda) \partial^- \phi(s_r|W\|\overline{W}))\right|
\le
\epsilon. \Label{19-6}
\end{align}
The relation (\ref{16-5}) also implies that
\begin{align}
0 \le &
\varphi(s_r-(1-\lambda)\delta|\lambda')
-\varphi(\overline{s}_r|\lambda')
\le
\varphi(s_r-(1-\lambda)\delta|\lambda')
-\varphi(s_r+ \lambda\delta|\lambda') \nonumber \\
\le &
[\epsilon-((\lambda \partial^+ \phi(s_r|W\|\overline{W})
+(1-\lambda) \partial^- \phi(s_r|W\|\overline{W}))]\delta
 \nonumber \\
\le &
(\epsilon -\partial^- \phi(s_r|W\|\overline{W}))
\delta
\Label{19-1}.
\end{align}
Since
\begin{align*}
\phi(s_r-(1-\lambda)\delta|W_{x^+}\|\overline{W}_{x^+})
\ge 
\phi(s_r+\lambda \delta|W_{x^+}\|\overline{W}_{x^+}),
\end{align*}
relations (\ref{16-2}) and (\ref{16-3}) guarantee that
\begin{align*}
0 \le &
\phi(s_r-(1-\lambda)\delta|W\|\overline{W})
-\phi(s_r-(1-\lambda)\delta|W_{x^+}\|\overline{W}_{x^+}) \\
\le &
\phi(s_r-(1-\lambda)\delta|W\|\overline{W})
-\phi(s_r+\lambda \delta|W\|\overline{W})
+\phi(s_r+\lambda \delta|W\|\overline{W})
-\phi(s_r+\lambda \delta|W_{x^+}\|\overline{W}_{x^+}) \\
\le &
(\epsilon -\partial^- \phi(s_r|W\|\overline{W}))
(s_r+\lambda \delta-\overline{s}_r)
+\delta \epsilon\\
\le &
(\epsilon -\partial^- \phi(s_r|W\|\overline{W}))
\delta
+ \delta \epsilon=
(2 \epsilon -\partial^- \phi(s_r|W\|\overline{W}))\delta.
\end{align*}
Therefore, 
\begin{align}
0 \le &
\phi(s_r-(1-\lambda)\delta|W\|\overline{W})
-\varphi(s_r-(1-\lambda)\delta|\lambda') \nonumber\\
\le &
\lambda'
(\phi(s_r-(1-\lambda)\delta|W\|\overline{W})
-
\phi(s_r-(1-\lambda)\delta|W_{x^+}\|\overline{W}_{x^+}))
+
(1-\lambda')
(\phi(s_r-(1-\lambda)\delta|W\|\overline{W})
-
\phi(s_r-(1-\lambda)\delta|W_{x^-}\|\overline{W}_{x^-})) \nonumber\\
\le &
\lambda' (\epsilon -\partial^- \phi(s_r|W\|\overline{W})) \delta
+(1-\lambda' )\delta \epsilon
\le
(\epsilon -\partial^- \phi(s_r|W\|\overline{W})) \delta
.\Label{19-2}
\end{align}
Since (\ref{16-2}) implies that
\begin{align*}
\phi(s_r-(1-\lambda)\delta|W\|\overline{W})-\phi(s_r|W\|\overline{W})
\le
(\epsilon -\partial^- \phi(s_r|W\|\overline{W}))
\delta,
\end{align*}
relations (\ref{19-1}) and (\ref{19-2}) guarantee that
\begin{align}
& |\varphi(\overline{s}_r|\lambda')-\phi(s_r|W\|\overline{W})| \nonumber\\ 
\le &
|\varphi(\overline{s}_r|\lambda')-\varphi(s_r-(1-\lambda)\delta|\lambda')|
+|\varphi(s_r-(1-\lambda)\delta|\lambda')-\phi(s_r-(1-\lambda)\delta|W\|\overline{W})|
+|\phi(s_r-(1-\lambda)\delta|W\|\overline{W})-\phi(s_r|W\|\overline{W})|
\nonumber\\
\le &
(4 \epsilon -3 \partial^- \phi(s_r|W\|\overline{W})) \delta
\le C_2 \delta, \Label{19-5}
\end{align}
where
\begin{align*}
C_2:=
4 -3 \partial^- \phi(s_r|W\|\overline{W}))
\ge 4 \epsilon -3 \partial^- \phi(s_r|W\|\overline{W}).
\end{align*}
Note that the constant $C_2$ does not depend on $\epsilon$ or $\delta$.

We choose a real number $\overline{r}:=
(1-\overline{s}_r)\varphi(\overline{s}_r|\lambda')+\varphi'(\overline{s}_r|\lambda')$.
Then, (\ref{19-5}), (\ref{19-6}), and the inequality $|s_r-\overline{s}_r|\le \delta$ imply that
\begin{align}
& |\overline{r}-r| \nonumber\\
\le &
|(1-\overline{s}_r)\varphi(\overline{s}_r|\lambda')-
(1-s_r)\phi(s_r|W\|\overline{W}))|
+|\varphi'(\overline{s}_r|\lambda')-
(\lambda \partial^+ \phi(s_r|W\|\overline{W})+(1-\lambda) \partial^- \phi(s_r+|W\|\overline{W}))|
\nonumber\\
\le &
|(1-\overline{s}_r)(\varphi(\overline{s}_r|\lambda')-\phi(s_r|W\|\overline{W}))|
+|\phi(s_r|W\|\overline{W})(s_r-\overline{s}_r)| 
+|\varphi'(\overline{s}_r|\lambda')-
(\lambda \partial^+ \phi(s_r|W\|\overline{W})+(1-\lambda) \partial^- \phi(s_r+|W\|\overline{W}))|
\nonumber\\
\le &
(1-\overline{s}_r) C_2 \delta
+|\phi(s_r|W\|\overline{W})|\delta 
+\epsilon
\le 
C_3 \delta +\epsilon,  \Label{19-7}
\end{align}
where
\begin{align*}
C_3 :=&
(2-s_r) C_2 + |\phi(s_r|W\|\overline{W})| \\
\ge &
(1-s_r+(1-\lambda)\delta) C_2 + |\phi(s_r|W\|\overline{W})| \\
\ge &
(1-\overline{s}_r) C_2 + |\phi(s_r|W\|\overline{W})| .
\end{align*}
Note that the constant $C_3$ does not depend on $\epsilon$ or $\delta$.
The function $\frac{-s\overline{r}- \varphi(s|\lambda')}{1-s}$
takes the maximum at $s=\overline{s}_r$.
Using (\ref{19-5}) and (\ref{19-7}), we can check that
this maximum is approximated by the value $\frac{-s_r r- \phi(s_r||W\|\overline{W})}{1-s_r}$
as
\begin{align}
& |\frac{-\overline{s}_r\overline{r}- \varphi(\overline{s}_r|\lambda')}{1-\overline{s}_r}
-\frac{-s_r r- \phi(s_r|W\|\overline{W})}{1-s_r}|\nonumber\\
\le &
|\frac{-\overline{s}_r\overline{r}- \varphi(\overline{s}_r|\lambda')}{1-\overline{s}_r}
-\frac{-s_r r- \phi(s_r|W\|\overline{W})}{1-\overline{s}_r}|
+
|\frac{-s_r r- \phi(s_r|W\|\overline{W})}{1-\overline{s}_r}
-\frac{-s_r r- \phi(s_r|W\|\overline{W})}{1-s_r}| \nonumber\\
\le &
|\frac{\overline{s}_r\overline{r}-s_r r}{1-\overline{s}_r}|
+|\frac{ \varphi(\overline{s}_r|\lambda')- \phi(s_r|W\|\overline{W})}{1-\overline{s}_r}|
+|\frac{-s_r r- \phi(s_r|W\|\overline{W})(s_r-\overline{s}_r)}{(1-\overline{s}_r)(1-s_r)}\nonumber\\
\le &
\frac{|(\overline{s}_r(\overline{r}-r)|+| r (\overline{s}_r-s_r)|}{1-\overline{s}_r}
+|\frac{ \varphi(\overline{s}_r|\lambda')- \phi(s_r|W\|\overline{W})}{1-\overline{s}_r}|
+|\frac{-s_r r- \phi(s_r|W\|\overline{W})}{(1-s_r+1)(1-s_r)}|\delta\nonumber\\
\le &
\frac{(-s_r+\delta)(C_3 \delta +\epsilon)+ r \delta}{2-s_r}
+|\frac{C_2 \epsilon}{2-s_r}|
+|\frac{-s_r r- \phi(s_r|W\|\overline{W})}{(2-s_r)(1-s_r)}|\delta \nonumber \\
\le & C_4 \epsilon + C_5 \delta,\Label{19-8}
\end{align}
where we choose $C_4$ and $C_5$ as follows.
\begin{align*}
C_4:=&
\frac{-s_r+1}{2-s_r}
+|\frac{C_2}{2-s_r}|\\
\ge & \frac{-s_r+\delta}{2-s_r}
+|\frac{C_2}{2-s_r}|\\
C_5:=&
\frac{(-s_r+1)C_3 + r \delta}{2-s_r}
+|\frac{-s_r r- \phi(s_r|W\|\overline{W})}{(2-s_r)(1-s_r)}|\\
\ge & \frac{(-s_r+\delta)C_3 + r \delta}{2-s_r}
+|\frac{-s_r r- \phi(s_r|W\|\overline{W})}{(2-s_r)(1-s_r)}|.
\end{align*}
Note that the constants $C_4$ and $C_5$ do not depend on $\delta$ or $\epsilon$.
Since
\begin{align*}
|\frac{-s r- \varphi(s|\lambda')}{1-s}
-\frac{-s \overline{r}- \varphi(s|\lambda')}{1-s}|
\le 
\frac{-s}{1-s}
|r-\overline{r}|
\le |r-\overline{r}|,
\end{align*}
(\ref{19-7}) implies that
\begin{align}
|\max_{s \le 0}\frac{-s r- \varphi(s|\lambda')}{1-s}
-\max_{s \le 0} \frac{-s \overline{r}- \varphi(s|\lambda')}{1-s}|
\le |r-\overline{r}|\le C_3 \delta +\epsilon.\Label{20-1}
\end{align}
Since $\varphi(s|\lambda')\le \phi(s|W\|\overline{W})$,
(\ref{20-1}) and (\ref{19-8}) guarantee that
\begin{align}
0\le 
\max_{s \le 0}\frac{-s r- \varphi(s|\lambda')}{1-s}
-
\frac{-s_r r- \phi(s_r|W\|\overline{W})}{1-s_r}
\le (C_4 +1)\epsilon + (C_3+C_5) \delta.\Label{20-2}
\end{align}
We define the distribution 
$P_{\lambda'} \in {\cal P}^2({\cal X})$ by
\begin{align*}
P_{\lambda'}(x^+)=\lambda', \quad 
P_{\lambda'}(x^-)=1-\lambda'.
\end{align*}
Since the function $x \to \log x$ is concave,
the inequality
\begin{align}
\varphi(s|\lambda')\le \phi(s|W\|\overline{W}|P_{\lambda'}) \Label{25-1}
\end{align}
holds.
Hence, (\ref{20-2}) and (\ref{25-1}) imply that 
\begin{align*}
0\le &
\inf_{P \in {\cal P}^2({\cal X})}
\max_{s \le 0}\frac{-s r- \phi(s|W\|\overline{W}|P)}{1-s}
-
\frac{-s_r r- \phi(s_r|W\|\overline{W})}{1-s_r} \\
\le &
\max_{s \le 0}\frac{-s r- \phi(s|W\|\overline{W}|P_{\lambda'})}{1-s}
-
\frac{-s_r r- \phi(s_r|W\|\overline{W})}{1-s_r}
\le (C_4 +1)\epsilon + (C_3+C_5) \delta.
\end{align*}
We take the limit $\delta \to +0$. After this limit, we take the limit $\epsilon \to +0$. Then, we obtain (\ref{16-7}).

Next, we prove the inequality (\ref{16-7}) when the maximum $\max_{s \le 0}
\frac{-s r- \phi(s|W\|\overline{W})}{1-s}$ does not exist.
The real number $R:=\lim_{s \to - \infty}\frac{\phi(s|W\|\overline{W})}{s}$
satisfies $r \ge - R$.
Thus,
\begin{align*}
\sup_{s \le 0} \frac{-s r- \phi(s_r|W\|\overline{W})}{1-s}
= r+R.
\end{align*}
For any $\epsilon >0$, there exists $s_0 <0$ such that
any $s < s_0$ satisfies that
\begin{align*}
R \le \frac{\phi(s_0|W\|\overline{W})-\phi(s|W\|\overline{W})}{s_0-s}
\le R+ \epsilon.
\end{align*}
We choose $x_0$ such that
\begin{align*}
\phi(s_0-1|W\|\overline{W})-\epsilon
\le
\phi(s_0-1|W_{x_0}\|\overline{W}_{x_0})
\le
\phi(s_0-1|W\|\overline{W}).
\end{align*}
Thus,
\begin{align*}
\phi(s_0|W_{x_0}\|\overline{W}_{x_0})
-\phi(s_0-1|W_{x_0}\|\overline{W}_{x_0})
\le
\phi(s_0|W\|\overline{W})
-\phi(s_0-1|W\|\overline{W})
+\epsilon
\le R + 2 \epsilon.
\end{align*}
Hence, for any $s < s_0$,
\begin{align*}
& \frac{\phi(s_0|W_{x_0}\|\overline{W}_{x_0})-\phi(s|W_{x_0}\|\overline{W}_{x_0})}{s_0-s} \\
\le &
\phi(s_0|W_{x_0}\|\overline{W}_{x_0})-\phi(s_0-1|W_{x_0}\|\overline{W}_{x_0}) \\
\le &
\phi(s_0|W\|\overline{W})
-\phi(s_0-1|W\|\overline{W})
+\epsilon
\le R + 2 \epsilon.
\end{align*}
Thus,
\begin{align*}
-r \le R \le 
\lim_{s \to -\infty}
\frac{\phi(s|W_{x_0}\|\overline{W}_{x_0})}{s}
\le R + 2 \epsilon.
\end{align*}
Therefore,
\begin{align*}
\sup_{s \le 0} \frac{-s r- \phi(s_r|W_{x_0}\|\overline{W}_{x_0})}{1-s}
\le r+R+ 2 \epsilon.
\end{align*}
Taking $\epsilon \to 0$, we obtain (\ref{16-7}).

\section{Concluding remarks and future study}
We have obtained a general asymptotic formula for the discrimination of two classical channels with adaptive improvement concerning the several asymptotic formulations.
We have proved that any adaptive method does not improve the asymptotic performance.
That is, the non-adaptive method attains the optimum performance in these asymptotic formulations.
Applying the obtained result to the discrimination of two quantum states by one-way LOCC, 
we have shown that one-way communication does not improve the asymptotic performance in these senses.

On the other hand, as shown in Section 3.5 of Hayashi\cite{Hayashi},
we cannot improve the asymptotic performance of
the Stein bound even if we extend the class of our measurement to the 
separable POVM in the $n$-partite system.
Hence, two-way LOCC does not improve the Stein bound.
However, other asymptotic performances in two-way LOCC and separable POVM have not been solved.
Therefore, it is an interesting problem to solve whether 
two-way LOCC improves the asymptotic performance for other than the Stein's bound.

Furthermore, the discrimination of two quantum channels (TP-CP maps) 
is an interesting related topic.
An open problem remains as to whether choosing input quantum states adaptively
improves the discrimination performance in an asymptotic framework.
The solution to this problem will be sought in a future study.

\section*{Acknowledgments}
The author would like to thank
Professor Emilio Bagan, Professor Ramon Munoz Tapia,
and Dr. John Calsamiglia for their interesting discussions.
The present study was supported in part by MEXT through a Grant-in-Aid for Scientific Research on Priority Area ``Deepening and Expansion of Statistical Mechanical Informatics (DEX-SMI),'' No. 18079014.

\bibliographystyle{IEEE}

\end{document}